\documentclass[
pre,
preprint,
]{revtex4-1}
\usepackage{graphicx}
\usepackage{amsmath}
\usepackage{amssymb}
\usepackage[light,math]{iwona}
\newcommand{\SUBEQ}[2][ ]{\begin{subequations} \label{#1} #2 \end{subequations}}
\newcommand{\FIGURE}[2][1]{\includegraphics[width=#1\textwidth]{#2}}

\newcommand{\Mstar}{M_{\mathrm{*}}}
\newcommand{\Msun}{M_{\mathrm{\odot}}}

\newcommand{\Rstar}{R_{\mathrm{*}}}
\newcommand{\kB}{K_{\mathrm{B}}}
\newcommand{\LEdd}{\mathcal{L}_\mathrm{Edd}}

\newcommand{\vS}{c_{\mathrm{S}}}
\newcommand{\charge}{\mathrm{e}}
\newcommand{\emass}{m_{\mathrm{e}}}
\newcommand{\pmass}{m_{\mathrm{p}}}
\renewcommand{\ne}{n_{\mathrm{e}}}

\newcommand{\km}{\, \mathrm{km}}
\newcommand{\ev}{\, \mathrm{eV}}
\newcommand{\kev}{\, \mathrm{keV}}
\newcommand{\Mev}{\, \mathrm{MeV}}
\newcommand{\cm}{\, \mathrm{cm}}
\newcommand{\second}{\, \mathrm{s}}
\newcommand{\gram}{\, \mathrm{g}}
\newcommand{\yr}{\, \mathrm{yr}}
\newcommand{\kelvin}{\, \mathrm{K}}
\newcommand{\gauss}{\, \mathrm{G}}
\newcommand{\erg}{\, \mathrm{erg}}

\newcommand{\visco}{\eta_{\mathrm{V}}}
\newcommand{\conduco}{\sigma_{\mathrm{B}}}

\newcommand{\<}[1]{\! \left( #1 \right)}
\newcommand{\virg}[1]{`#1'}

\newcommand{\Eref}[1]{Equation \eqref{#1}}
\newcommand{\Sref}[1]{Section \ref{#1}}
\newcommand{\Fref}[1]{Figure \ref{#1}}

\newcommand{\PRD}{Phys. Rev. D}
\newcommand{\PRE}{Phys. Rev. E}
\newcommand{\APJ}{Astrophys. Journal}
\newcommand{\apjl}{Astrophys. Journal Lett.}

\newcommand{\mnras}{Mon. Not. R. Astron. Soc.}
\newcommand{\aap}{Astron. Astrophys.}
\newcommand{\NAT}{Nature Phys. Science}

\newcommand{\aj}{Astron. Journal}

\newcommand{\NAR}{New Astron. Rev.}

\newcommand{\SA}{Sov. Astron.}

\newcommand{\pasp}{Publ. Astron. Soc. Pac.}
\newcommand{\pasj}{Publ. Astron. Soc. Jap.}
\newcommand{\RMP}{Rev.s of Modern Phys.}

\begin{document}

\title{
Plasma Phenomenology in Astrophysical Systems: Radio-Sources and Jets
}

\author{Giovanni Montani}
\email{giovanni.montani@frascati.enea.it} 
\affiliation{ENEA -- C.R. Frascati, U.T. Fus. (FUSMAG Lab) -- Via Enrico Fermi 45, (00044) Frascati (RM), Italy}
\affiliation{Physics Department, `Sapienza' University of Rome -- Piazzale Aldo Moro 5, (00185) Roma, Italy}

\author{Jacopo Petitta}
\email{petitta.jacopo@gmail.com}
\affiliation{Physics Department, `Sapienza' University of Rome -- Piazzale Aldo Moro 5, (00185) Roma, Italy}

\begin{abstract}
We review the plasma phenomenology in the astrophysical sources which show appreciable radio emissions, namely Radio-Jets from Pulsars, Microquasars, Quasars and Radio-Active Galaxies.
A description of their basic features is presented, then we discuss in some details the links between their morphology and the mechanisms that lead to the different radio-emissions, investigating especially the role played by the plasma configurations surrounding compact objects (Neutron Stars, Black Holes).
For the sake of completeness, we briefly mention observational techniques and detectors, whose structure set them apart from other astrophysical instruments.
The fundamental ideas concerning Angular Momentum Transport across plasma accretion disks -- together with the disk-source-jet coupling problem -- are discussed, by stressing their successes and their shortcomings.
An alternative scenario is then inferred, based on a parallelism between astrophysical and laboratory plasma configurations, where small-scale structures can be found.
We will focus our attention on the morphology of the radio-jets, on their coupling with the accretion disks and on the possible triggering phenomena, viewed as profiles of plasma instabilities.
\end{abstract}

\maketitle

\section{Introduction}
The branch of astronomy that investigates celestial objects in the range of radio frequencies is called Radio Astronomy.
It is born in the 1930s, when the first detection of radio waves from an astronomical object led to the discovery of the radio source Sagittarius A in the densest part of the Milky Way.
The polarization and the frequency range led Jansky \cite{jansky33} to rule out thermal emission from galactic gas and dust and ascribe the emission to free electrons embedded in a strong magnetic field, originated from the complex of objects found in the neighbourhood of the galactic center, and with the contribution of the Super-Massive Black Hole (SMBH) named Sagittarius A*.
From that time on, observations have identified some different sources responsible for radio emission, including Nebulae, Intergalactic and Interstellar Medium, as well as brand new classes of objects, such as Radio Galaxies, Active Galactic Nuclei (AGN), Quasars and Pulsars.
The emission from this new sources is totally non-thermal and often found coming from jets and jet-like structures, so that it is intrinsically different from the thermal radiation revealed in the cold and diffuse objects listed before, which accurately reproduces blackbody spectra peaked around 60 GHz and corresponding to source temperatures of the order of the unity of K.
Although it is beyond the scope of this work, the discovery of the Cosmic Microwave Background has also to be ascribed in the merits of radio astronomy: in the 1960s, Penzias \& Wilson \cite{PW65} laid the foundations for the field of experimental cosmology, measuring temperature excesses on a radio antenna and providing compelling evidence for the Big Bang.

The observation of intense radio sources 
allowed to characterize the morphology of 
compact objects and their capability to 
generate highly collimated jets. 
The physics governing 
the radio-emission scenario is directly 
linked to the accretion phenomena, especially for what concerns the 
jet formation, as a result of unstable modes 
taking place in the plasma equilibrium configuration of 
the accreting profiles.
In this paper we will describe the peculiarities and common features of astrophysical radio sources, in order to highlight the relevance of plasma configurations and instabilities in the triggering of jets, which turn out to be ultimately responsible for radio emission in the examined objects.
We will also investigate the basics of accretion phenomena, which are intimately coupled with the formation and launch of radio-jets.

The paper is organized as follows.
In \Sref{sec:sources}, we briefly review the principal astrophysical systems active in the radio band of electromagnetic spectrum.
In \Sref{sec:detectors}, the basic information about the detectors used in radio astronomy are given.
In \Sref{sec:emission}, we explain the main processes responsible for the generation of radio waves in astrophysical settings.
In \Sref{sec:accretion}, the physics of accretion and Angular Momentum Transport (AMT) is depicted, in order to trace down the parallelism between plasma astrophysics and laboratory plasma physics, and a new model for jet-triggering is shown.
Concluding remarks will follow in \Sref{sec:conclusions}.

\section{Radio Sources} \label{sec:sources}
\begin{figure} 
\centering
\FIGURE[.48]{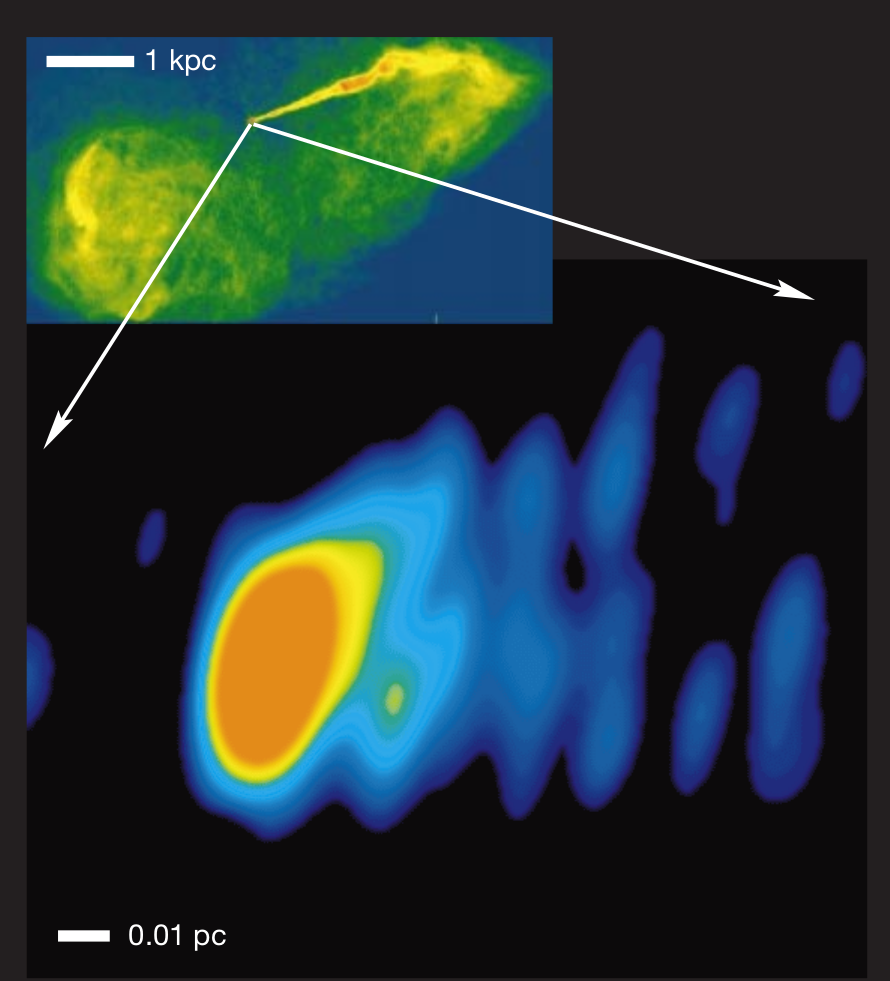}
\caption{ \label{fig:m87}
(Color online)
The supergiant elliptical galaxy M87 (also identified as Virgo A or NGC 4486) offers clear views of the large-scale jet-lobe structure and of the central SMBH at the same time.
\emph{Top left:} 15-GHz VLA image illustrating the jet, going from the nucleus to the diffuse lobe.
\emph{Bottom:} Pseudo-colour rendition of the nucleus of M87 at 43 GHz. 
Courtesy of W. Junor (Los Alamos National Laboratory), J. A. Biretta \& M. Livio (Space Telescope Science Institute).
Reprinted by permission from Macmillan Publishers Ltd: [Nature] \cite{junor99}, copyright (1999).
}
\end{figure}

\paragraph{ \bf Radio Galaxies (Prototypes: Centaurus A, Messier 87):}
Radio-loud galaxies are active galaxies with luminosities up to $10^{39}$ W between 10 MHz and 100 GHz.
Such massive radio emission is mostly due to the synchrotron process, as inferred from its very smooth, broad-band nature and strong polarization.
This implies that the radio-emitting plasma contains relativistic electrons (with Lorentz factors $\gamma \sim 10^4$) embedded in significant magnetic fields (with strength $B \sim 10^{-5} \gauss$).
The radiation displays a wide range of structures in radio maps, the most common being called lobes: double, often fairly symmetrical, roughly ellipsoidal structures placed on either side of the active galactic nucleus.
A significant minority of low-luminosity sources exhibit structures usually known as plumes which are much more elongated, yet characterized by the same kind of emission.
Some radio galaxies show one or two long narrow features known as jets coming directly from the nucleus and going to the lobes (the iconic example of M87 is shown in \Fref{fig:m87}); in this case, the host galaxies are almost exclusively large elliptical galaxies.

Observed emission comes from the interaction between the axial jets and the intergalactic medium, and is sensitively modified by relativistic beaming of the emitted photons, which lead to the Fanaroff-Riley classification:
\emph{FR-I} sources are brightest towards the centre, with bright jets which radiate a significant amount of their energy away as they travel and are decelerated to sub-relativistic speeds by interaction with the external medium -- they are low-luminosity sources; 
\emph{FR-II} sources are brightest at the edges, with faint but highly relativistic jets and bright radio hot-spots which shows how energy is efficiently transported to the end of the lobes -- they are high-luminosity sources.
Radio-loud active galaxies are interesting also because they can be detected and identified at large distances, making them valuable tools for observational cosmology.
A useful review on this matters is provided by Urry \& Padovani \cite{urry95}, and detailed informations on M87 -- the most famous object in this class -- can be found in Biretta \emph{et al.} \cite{biretta91}.

\paragraph{\label{par:quasar} \bf Quasars and Blazars (Prototypes: Virgo 3C 273, BL Lacertae):} 
A QUASi-stellAR radio source is a very distant active galactic nucleus, point-like but still extremely luminous.
Resolving the actual structure of these objects is forbidden for most are farther than $3 \times 10^9$ light-years, but their visibility is assured by brightnesses of the order of $10^{40}$ W, roughly equivalent to $2\times 10^{12}$ sun-like stars.
Quasars were first thought simply as high-redshift sources of electromagnetic energy, including radio waves and visible light; they are now identified as compact regions in the center of massive galaxies surrounding SMBHs.
Their size can be $10 \div 10^4$ times the Schwarzschild radius of the BH.
The energy emission is powered by accretion (see below, \Sref{sec:accretion}), since the matter located outside the event horizon suffer huge gravitational stresses and immense friction.
Accreting matter is unlikely to fall directly in, but will have to get rid of its angular momentum, resulting in the formation of a plasma accretion disk revolving around the central BH.
The actual accretion rate needed to explain the high luminosities observed ranges from 10 to $30 \, \Msun / \yr$, where $\Msun \simeq 1.99 \times 10^{33} \gram$ is the solar mass.
Quasars may also be ignited or re-ignited from normal galaxies when infused with a fresh source of matter: an actual new Quasar could form from the collision of the Andromeda Galaxy with our own Milky Way galaxy, an event expected in approximately $3 \div 5$ billion years.
\begin{figure*} 
\centering
\includegraphics[width=\textwidth]{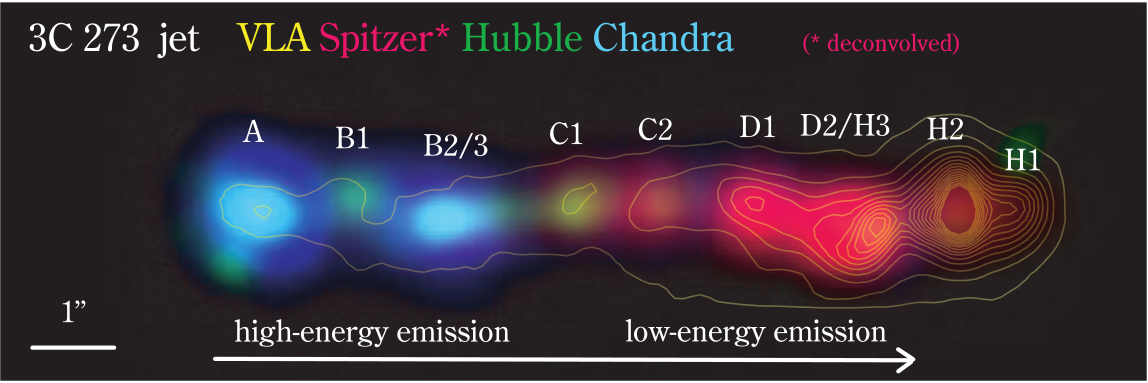}
\caption{\label{fig:jet}
(Color online) Composite image of the jet in Virgo 3C 273.
The data are coded as follows: Spitzer \virg{deconvolved} @3.6 $\mu$m (in red), Hubble Space Telescope \virg{UV excess} (in green), Chandra @(0.4 -- 6) keV (in cyan).
The IR contribution peaks towards the head of the jet, on the D1-2 knots; the UV/optical emission is strongly localized but it is founded all along the structure, peaking on all the knots numbered with 1; the X-rays come mainly from the accretion disk, so their peak is located in the surroundings of the SMBH and near the base of the jet, on the A knot.
The VLA 2-cm Radio contours are superimposed on the image, with the strongest radio source identified as H2, usually called `radio hot-spot'.
Reproduced by permission of the American Astronomical Society \cite{uchiyama06}.
}
\end{figure*}

Although Quasars can be detected over the entire observable spectrum, the strong radio emission originates again from highly-relativistic electrons in jets (see \Fref{fig:jet}, where the different bands dominating the emission are highlighted along the jet).
In particular, when the jet points towards the observer, a Quasar appears as a Blazar (named after the prototype object BL Lacertae): in this case the observer finds the highest luminosities, the most rapid variabilities and the highest degree of polarization.
Blazars often show regions experiencing apparently superluminal expansion within the first few parsecs of their jets, probably due to relativistic shock fronts which suffer relativistic aberration.
This phenomenon only occurs in this class because of the specific coincidence of electrons and photons travelling at ultra-relativistic speeds along the line of sight which joins the source and the observer.
At every point of their path the high-velocity jets are emitting photons, which do not approach the observer much more quickly than the jet itself.
Light emitted over hundreds of years of travel thus arrives at the observer over a much smaller time period (about ten years), giving the illusion of superluminal motion without any violation of special relativity.

\paragraph{\bf Supernova Remnants (Prototypes: Crab Nebula, Cassiopeia A):}
A SuperNova Remnant (SNR) is the structure resulting from the supernova-explosion of a star, which leaves a collapsed core in the form of a Neutron Star (NS) or a Black Hole (BH), depending on the mass of the original star.
This compact object is surrounded by the ejected material expanding from the explosion, which is bounded by a shock wave sweeping the interstellar medium along the way.
All these features make up the whole SNR and contribute to its composite spectrum.

\begin{figure*}
\centering
\FIGURE[.49]{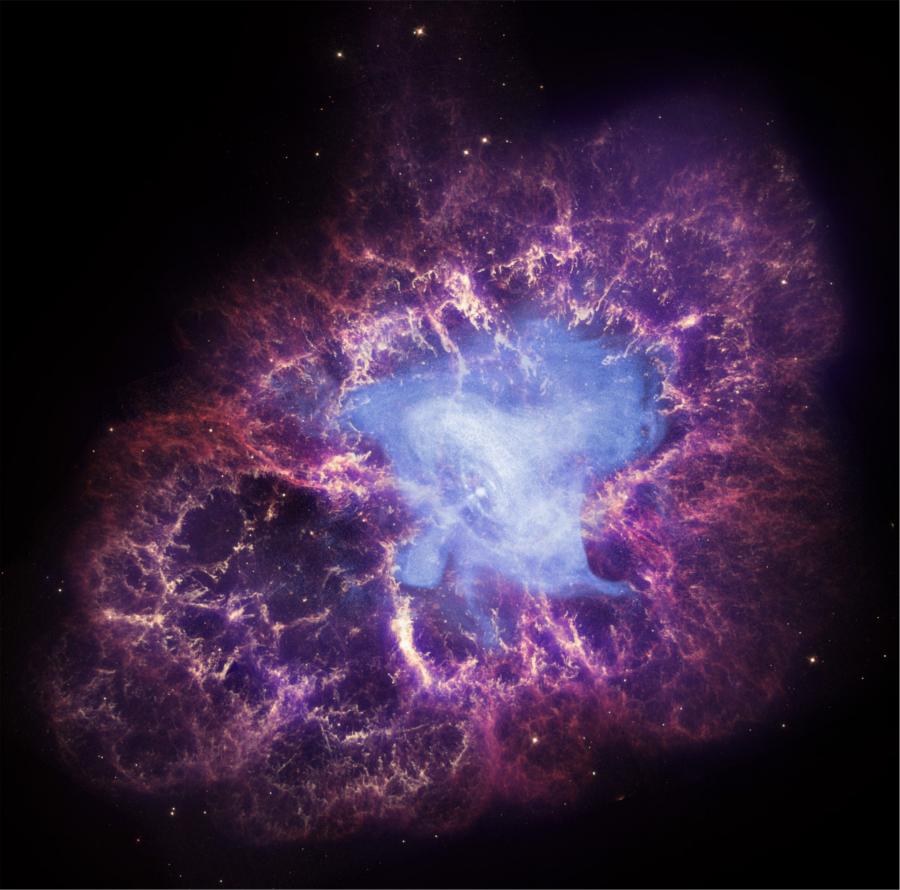}
\FIGURE[.49]{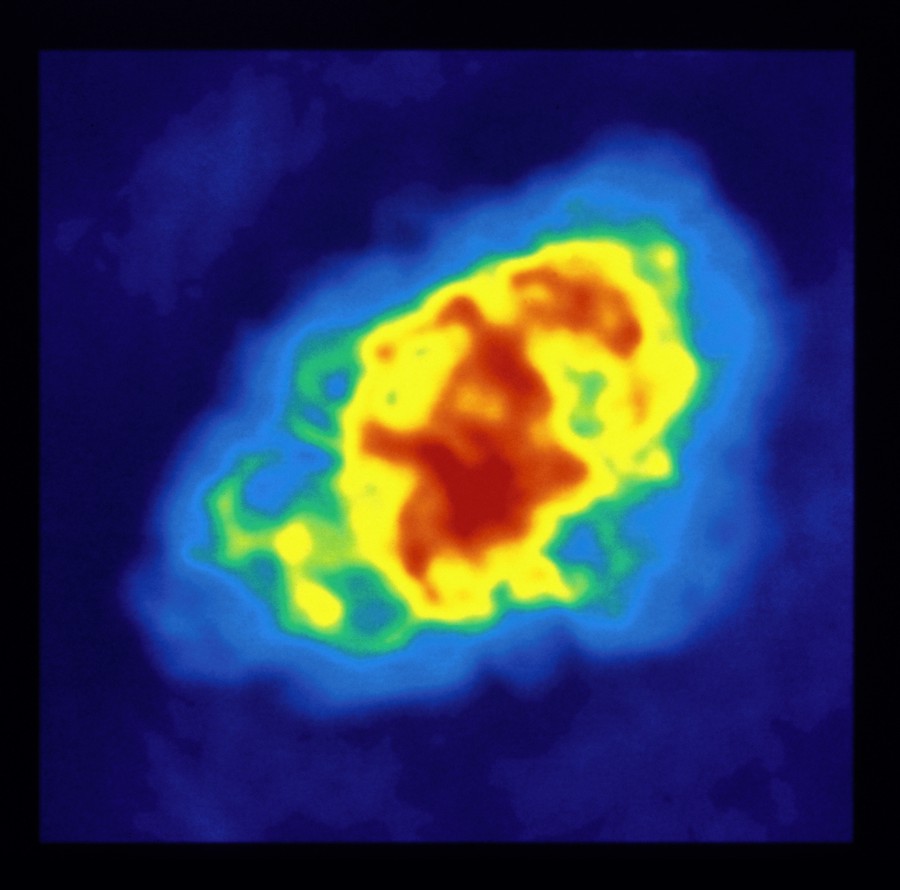}
\caption{ \label{fig:crab}
(Color online)
Composite image of the Crab Nebula (designated also as M1 or NGC 1952), highlighting the different bands of the detected spectrum, corresponding to different regions of the SNR structure.
\emph{Left:} The Chandra X-ray image is shown in blue, the Hubble Space Telescope optical image is in red and yellow, and the Spitzer Space Telescope's infrared image is in purple. All credits to NASA, specifically: NASA/CXC/J.Hester (ASU) for X-Ray; NASA/ESA/J. Hester \& A. Loll (ASU) for optical; NASA/JPL-Caltech/R. Gehrz (Minnesota Univ.) for IR.
\emph{Right:} The diffuse thermal radio emission measured by VLA via interferometry, peaked in the neighbourhood of the central pulsar. Image courtesy of NRAO/AUI/NSF.
}
\end{figure*}

An SNR passes through different stages as it expands, corresponding to different emissions; \Fref{fig:crab} shows the well-documented case of the Crab Nebula.
The expanding layer of shocked circumstellar and interstellar gas produces strong X-ray emission; then it starts cooling to form a thin (less than 1 pc), dense shell surrounding the central NS or BH.
The shell can be clearly seen in the optical portion of the spectrum, as the radiation is emitted from recombining ionized hydrogen and oxygen atoms.
As the shell continues to expand by virtue of its own momentum, the inner layer continues to cool: the dominating emission is now radio, from neutral hydrogen atoms.
This radiation is fairly thermal, with a blackbody temperature $T_\mathrm{bb} \sim 30 \kelvin$, and it is not due to plasma jets.
A review of models and observations in this field is offered by Asvarov \cite{asvarov06}.
In the peculiar case of the Crab Nebula, this well-grounded picture has been challenged by (or, at least, integrated with) the emission of non-periodic highly energetic $\gamma$-flares, recently discovered by AGILE \cite{flareAGILE} and confirmed by FERMI \cite{flareFERMI}.

\paragraph{\bf Radio Pulsars (Prototype: Crab Pulsar PSR B0531+21):}
A PULSating stAR is a highly magnetized, rotating NS that emits a collimated beam of electromagnetic radiation, which can only be observed when pointing toward the Earth.
A very precise interval between pulses can be identified, ranging from milliseconds to seconds for an individual Pulsar, thanks to the short and regular rotational period that a NS usually exhibits.
The magnetic axis of the Pulsar determines the direction of the beam, and the misalignment with the rotational axis causes the beam to be seen once for every rotation of the NS, in a fashion that goes under the name of lighthouse effect.
The beam exploits the rotational energy of the NS, which generates an electrical field from the movement of the very strong magnetic field, resulting in the acceleration of protons and electrons on the star surface.
The particles turn out collimated in a plasma beam emanating from the poles of the magnetic field.
The rotation slows down and the period becomes longer as electromagnetic power is lost via radiative emission.
For further informations, we refer to the detailed review  \cite{beskin99}.

Rotating Radio Transients (RRATs) are sources of short and moderately bright radio pulses, which were first discovered \cite{RRAT} in 2006, and are observed only in the radio band.
RRATs are thought to be special Pulsars, i.e. again rotating magnetized NSs but emitting more sporadically and/or with higher pulse-to-pulse variability than the bulk of the known Pulsars, which are usually very regular.
The time intervals between detected bursts range from seconds to hours, thus radio emission from RRATs is typically only detectable for less than one second per day.
Currently there is no complete model explaining the irregularity of RRAT pulses \cite{RRAT12}.

\paragraph{\bf Microquasars (Prototype: Cygnus X-1):}
A Microquasar is a radio emitting X-ray binary system.
They are named after Quasars because of some common characteristics: strong and variable radio emission resolvable as a pair of radio-jets, and an accretion disk surrounding a compact object which is either a NS or a BH.
In Microquasars, the mass of the compact object is only a few solar masses, the accreted mass comes from a normal star, and the accretion disk is very luminous in the optical and X-ray regions. 
It is worth noting that the variability time-scales are proportional to the mass of the compact object, therefore a Microquasar may show in one day what ordinary Quasars take centuries to go through, because of the SMBH accretor in the latter case.

Radio Pulsars (rotation-powered isolated Pulsars) and Microquasars (accretion-powered Pulsars in binary systems) exhibit very different spin behaviours although it is accepted that both kinds are manifestations of rotating magnetized NSs.
The major differences are as follows.
Radio Pulsars have periods on the order of milliseconds to seconds, and all radio Pulsars are losing angular momentum and slowing down; in contrast, Microquasars display a variety of spin behaviours: some are observed to be continuously spinning faster or slower (with occasional reversals in these trends), while others show either little change in pulse period or display random spin-down/spin-up behaviour.
The difference is rooted in the physical nature of those two Pulsar classes: almost all Radio Pulsars are single objects which are radiating away their rotational energy; Microquasars are, instead, members of binary star systems which accrete matter from either stellar winds or accretion disks.
The coupling of AMT between disk and NS may cause the spin rate to increase or decrease at rates that are often hundreds of times faster than the typical spin-down rate in Radio Pulsars; here also lies the fingerprint of a crucial relation between accretion processes, disk structure, AMT and jet triggering.
An up-to-date discussion and further details of the relevance of this kind of systems are in Gallo \cite{gallo10}.

\setcounter{paragraph}{0}
\section{Radio Detectors} \label{sec:detectors}
\begin{figure*}
\centering
\FIGURE{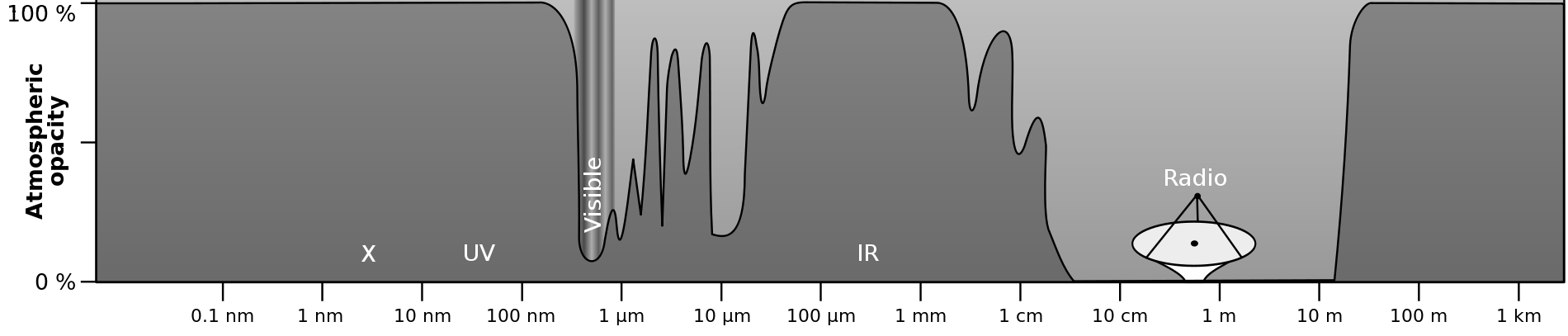}
\caption{ \label{fig:window}
Plot of atmospheric opacity versus radiation wavelength, showing the optical and radio windows where opacity falls to zero (or transmittance grows to one). $\gamma$, X and UV light (under 300 nm) are absorbed mainly by $O_2$ and $O_3$, while different molecular compounds ($H_2 O$ among others) block IR radiation. Wavelengths over 10 m are reflected by the ionosphere. Image adapted from an online published image; all credits to NASA.
}
\end{figure*}
The main research instruments in radio astronomy are large antennas referred to as radio telescopes, used singularly or in an array -- in which case, supplementary techniques like radio interferometry and aperture synthesis are adopted.
Observations from the Earth's surface are limited to wavelengths that can pass through the ionosphere, which reflects waves with frequencies less than its characteristic plasma frequency (i.e., wavelengths longer than $\sim 10$ m), while water vapour absorption interferes at higher frequencies.
This defines a \virg{radio window} which spans wavelengths from centimeters to tens of meters and is three orders of magnitude wider than its optical and infrared counterparts, as it is shown in \Fref{fig:window}.
Due to this huge range, radio telescopes vary in design, size, and configuration: instruments operating at less than 30 cm (above 1 GHz) range in size from 3 to 90 m; telescopes working at wavelengths from 30 cm to 3 m (100 MHz $\div$ 1 GHz) are usually well over 100 meters in diameter.
At these wavelengths \virg{dish} style radio telescopes predominate, with angular resolution determined by the diameter of the dish.
At wavelengths between 3 and 30 m ($10 \div 100$ MHz), they are either directional antennas or large stationary reflectors with moveable focal points, whose reflector surfaces can be constructed from coarse wire mesh.

Beyond single telescopes, radio interferometers consist of arrays of radio telescopes, widely separated but usually connected using some type of transmission line.
Interferometry can also be carried out by independent recording of the signals at the various antennas, and later correlating the recordings in a process known as Very Long Baseline Interferometry.
This slightly increases the total signal collected, but its primary purpose is to vastly increase the resolution through the aperture synthesis working by means of the superposition principle, which eventually creates a combined telescope equivalent in resolution (though not in sensitivity) to a single antenna whose diameter is equal to the spacing of the antennas furthest apart in the array.
The drawback is that it does not collect as many photons as a large instrument of that size: thus it is mainly useful for fine resolution of the more powerful astronomical sources.

In the long list of relevant radio telescopes and interferometers, it is worth mentioning:
the \emph{Arecibo Radio Telescope} (Puerto Rico), whose 305-m dish is a coated hollow in the ground;
the \emph{Very Large Array} at Socorro (New Mexico, USA), which has 27 telescopes -- each with a 25-m dish -- and 351 independent baselines at once, arrayed along the three 21-km arms of a Y-shape;
the \emph{Low-Frequency Array (LOFAR)}, operated by the Netherlands Institute for Radio Astronomy, which at the present time is the largest connected radio telescope, based on a vast array of about 25000 omni-directional dipole antennas (concentrated in 48 larger stations across the Northern Europe), and can achieve a total effective area up to 0.3 km$^2$.

\section{Radio Emissions} \label{sec:emission}
\paragraph{\bf Synchrotron Radiation,}
generated by astronomical objects because of relativistic electrons spiralling through magnetic fields, is the main source of detected non-thermal radio waves.
It features broad-band power-law spectra and a strong degree of polarization.
For non-relativistic motion (cyclotron radiation), the radiation spectrum shows a main spike at the fundamental frequency corresponding to the orbital motion, called the gyration frequency $\nu_\mathrm{C} = \charge B / \emass c$, where $\charge$ and $\emass$ are electronic charge and mass respectively.
But for this emission to be strong enough to have any astronomical significance, the electrons must be travelling at nearly the speed of light, i.e. with high Lorentz factor $\gamma \gg 1$; in this case, the radiation is compressed into a small range of angles $\sim \gamma^{-1}$ around the instantaneous velocity vector of the particle.
This is called \virg{beaming}, and it results in a spreading of the energy spectrum depending on the transverse momentum of the particle.
There is a maximum photon energy that can be radiated, which is proportional to the field strength and inversely proportional to the particle momentum.
Since this usually has a power-law distribution, synchrotron spectra have a power-law shape with measurable spectral index $s$, such that the flux is $F\<{\nu} \propto \nu^{-s}$ (mind that a different convention exists, with the opposite sign for $s$).
This kind of radiation is commonly detected in the radio region of the spectrum, although it extends to the X-ray band and beyond.
It is worth noting that as the electron travels along a magnetic field line and emits photons, it gives up energy and the more energy it loses, the wider becomes the trajectory, so that radiation is emitted at a longer wavelength, eventually peaking only on the radio range.

Synchrotron radio emission was first identified in the M87 jet \cite{burbidge56}.
Such jets have been confirmed by the Hubble telescope as apparently superluminal, travelling at $6c$ as seen from our planetary frame.

\begin{figure} 
\centering
\FIGURE[.48]{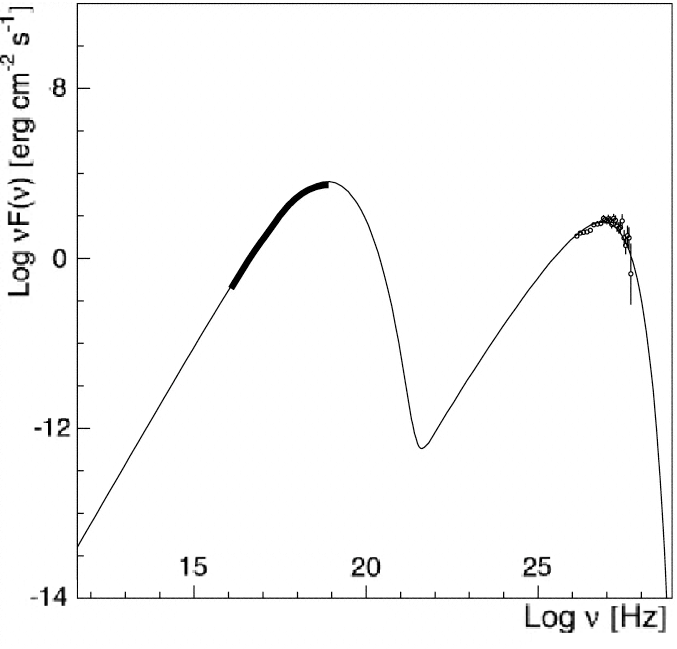}
\caption{ \label{fig:spectrum}
The pure synchrotron (peak near $10^{19}$ Hz) and synchrotron self-Compton (peak near $10^{27}$ Hz) spectra of the Blazar Markarian 501 (UGC 10599), including data and best fit, expressed as flux density per logarithmic frequency range. It is worth noting the difference in the spectral indexes of the two peaks, with a softer $s \simeq 2.2$ for the synchrotron and a harder $s \simeq 2.7$ for the self-Compton \cite{mkn11}.
Adapted from Figure 7 in Konopelko \emph{et al.} \cite{konopelko03}; reproduced by permission of the American Astronomical Society.}
\end{figure}

\paragraph{\bf Inverse Compton Scattering} is a combined effect of synchrotron emission and electron-photon scattering.
Although the total radiation field is fairly isotropic in the rest frame of the source, it is extremely anisotropic when looking at the individual ultra-relativistic electrons producing the synchrotron radiation: relativistic aberration causes nearly all ambient photons to be emitted within an angle $\gamma^{-1}$.
Thomson scattering of this highly anisotropic radiation systematically reduces the electron kinetic energy and converts it into inverse-Compton (IC) radiation by up-scattering radio photons to become optical or X-ray photons.
E.g., isotropic radio photons emitted at $\nu_0=1$ GHz, IC-scattered by electrons having $\gamma = 10^4$, will be up-scattered to the average frequency $\langle \nu \rangle = 4 \gamma^2 \nu_0 / 3 \simeq 1.3 \times 10^{17}$ Hz, corresponding to X-ray radiation.
Self-Compton radiation results from synchrotron radiation IC-scattered by the same relativistic electrons that firstly produced it.
This critical feedback is very sensitive to the source brightness temperature, so IC losses cool the relativistic electrons very efficiently and very rapidly if the brightness temperature exceeds $ T_\mathrm{b} \simeq 10^{12} \kelvin$ in the rest frame of the source.
Radio sources with brightness temperatures significantly high in the observer's frame are either Doppler boosted or not incoherent synchrotron sources (e.g., pulsars are coherent radio sources).
The active galaxy Markarian 501 (see \Fref{fig:spectrum}) emits strong synchrotron self-Compton radiation and the radio emission approaches this rest-frame brightness limit for incoherent synchrotron radiation.

\paragraph{\bf Spin-Flip Radiation,} identified as the 21-cm line or HI line, refers to the spectral line created by a change in the energy state of neutral hydrogen atoms: specifically, the atomic transition between the two hyperfine levels of the hydrogen $\phantom{.}^2$S ground state (with an energy difference of $5.87433 \, \mathrm{\mu eV}$, corresponding to a wavelength of $21.106 \cm$ in free space).
The hyperfine splitting is due to the magnetic coupling between electron and proton spins: the lowest-energy configuration arises in the anti-parallel spin (parallel magnetic moments) case, as an inherently quantum-mechanical result, against the predictions of classical mechanics.
The transition between those splitted levels is highly forbidden -- it is only allowed by quadrupole interaction or random low-energy collisions -- so that it has an extremely small probability and consequently a very long lifetime of $3.4 \times 10^{14} \second$ ($10^7$ years); this is the average time needed by a single isolated atom to undergo this transition spontaneously.
Although it is unlikely to be seen in a laboratory on Earth, this emission line is easily observed by radio telescopes, as the total number of atoms of neutral hydrogen available along any line-of-sight in the interstellar medium is very large.
Detection is also eased since those radio waves can pass through the interstellar dust -- that is opaque to visible light -- because of a wavelength much greater than the typical dimension of grains.
It is also worth noting that the line has an extremely small natural width because of its long lifetime, so most broadening is due to cumulative Doppler shifts caused by the motion relative to the observer.
Assuming that the hydrogen atoms are uniformly distributed throughout a galaxy, each line of sight will reveal a hydrogen line, with the only differences between each of these lines due to the different Doppler shifts.
Hence, one can calculate the relative speed of each arm of a galaxy and trace back the rotation curve, leading to indirect estimations of the mass of the galaxy and to the first evidence for the presence of dark matter \cite{HII70}.
Hydrogen line observations have also been used to put limits on any changes over time of the universal gravitational constant and to study dynamics of individual galaxies.

\setcounter{paragraph}{0}
\section{Accretion Power} \label{sec:accretion}
The extraction of gravitational potential energy is the principal source of power in several types of close binary systems, and is believed to provide the power supply in AGN and Quasars; it is in particular way more efficient than the nuclear fusion which fuels the stars \cite{FKR02}.
A body of mass $\Mstar$ and radius $\Rstar$, accreting a mass $m$ onto its surface, can release an amount of potential energy given by
$
\Delta E_\mathrm{grav} = {G \Mstar m} / {\Rstar}
$,
which is about $10^{20} \erg / \gram$ for a typical NS with $\Mstar \sim \Msun$ and $\Rstar \sim 10 \km$.
Let us compare this with the energy extracted -- by nuclear fusion reactions with maximum efficiency -- from a mass $m$ of hydrogen burning into helium, giving a release of
$
\Delta E_\mathrm{nuc} = 0.007 m c^2 \, ,
$
which is about $6 \times 10^{18} \erg / \gram \simeq {\Delta E_\mathrm{grav}}/{20}$.
The relative efficiency of the two mechanisms is determined by the compactness $\Mstar / \Rstar$ of the accreting object, favouring NSs and BHs.
Nuclear burning is more efficient for White Dwarfs (WDs), yet these objects rapidly run out of available nuclear fuel after short-duration bright events known as Nova outbursts.
Thus they fall back upon the accretion as the main source of power.
For a fixed value of the compactness, the total luminosity of an accreting system depends on the accretion rate $\dot{m}$, which may itself be determined by the momentum transferred from the emitted radiation to the accreting material via scattering and absorption, eventually leading to the existence of a maximum luminosity.

The net inward force on an electron-proton pair vanishes at the Eddington luminosity:
\begin{equation} \label{eq:Edd}
\LEdd = \frac{4 \pi G \Mstar \pmass c}{\sigma_\mathrm{T}} \simeq {1.3 \times 10^{38}} \frac{\Mstar}{\Msun} \erg / \second \, ,
\end{equation}
such that at greater luminosities the accretion would be halted because of unbalanced outward radiation pressure ($\sigma_\mathrm{T}$ is the Thomson scattering cross-section).
The Eddington limit estimation yields some other arguments:
for normal stars with a given $\Mstar-\mathcal{L}$ relation, we get a maximum stable mass;
for accretion powered objects we get an upper bound on the steady accretion rate;
last, it strongly suggests that AGN power supplies have to be SMBHs. 
If the nuclear burning is their main source, AGN/Quasars should require a huge accretion rate exceeding $\dot{m} \gtrsim 250 \Msun / \yr$ to reach the observed $\mathcal{L} \gtrsim 10^{47} \erg / \second$.
Otherwise, setting $\Rstar$ at the Schwarzschild radius and assuming a $10\%$ conversion efficiency for the accretion power, we get:
\begin{equation}
W_\mathrm{grav} \simeq 0.1 \dot{m} c^2 \gg W_\mathrm{nuc} \simeq 0.007 \dot{m} c^2 \, ,
\end{equation}
and a reasonable $\dot{m} \sim 20 \Msun / \yr$.
To efficiently radiate this power at less than the Eddington limit, a mass of the central object exceeding $10^9 \Msun$ is required by Equation \eqref{eq:Edd}, identifying an accreting SMBH.

\paragraph{\bf Spectra and jet-disk coupling --}
We can expect reliable ranges for emission frequencies of accretion-powered systems by means of two limits, for given $\Mstar$ and $\Rstar$.
If the given power $\mathcal{L} \simeq \LEdd$ is radiated as a blackbody spectrum we can infer a lower bound $T_\mathrm{bb}$ (no source can radiate a given flux at less than the blackbody temperature), while if the gravitational potential is turned entirely into thermal energy we can infer an upper bound $T_\mathrm{th}$.
Characterizing the continuum spectrum by a temperature $T_\mathrm{rad}$, we can write: 
\begin{equation}
T_\mathrm{bb}^4 \simeq \dfrac{\LEdd}{\sigma_\mathrm{S} 4 \pi \Rstar^2} 
\xleftarrow[\text{thick}]{\text{opt.}}
T_\mathrm{rad}
\xrightarrow[\mathrm{thin}]{\mathrm{opt.}}
T_\mathrm{th} \simeq \dfrac{G \Mstar \pmass}{\kB \Rstar} \, ,
\end{equation}
depending on the opacity of the accreting material ($\sigma_\mathrm{S}$ is the Stefann's constant). 
For a few-solar-masses NS or BH, we expect photon energies in $1 \kev \div 50 \Mev$, qualifying the systems as medium X-ray emitters up to $\gamma$-ray sources; for a WD we obtain $6 \ev \div 100 \kev$, revealing optical, UV and soft X-ray sources.
In both cases the radio emission comes from the jet ejection, while the explained accretion-related emission is produced at the core of the system and often due to an accretion disk.

Bipolar jets and accretion disks are tightly associated in both Quasars and Microquasars, where magnetohydrodynamic processes let them get rid of rotational energy through the poles by means of matter/energy jets, while the bulk can fall onto the gravitational attractor.
This accretion-ejection coupling around compact objects needs time intervals longer than years for SMBHs in Quasars, but it has been observed on time-scales shorter than an hour on Microquasars, e.g. the IR/Radio synchrotron flares in GRS 1915+105 \cite{mirabel98}.
A sudden fall in X- and $\gamma$-ray luminosity, rapidly followed by peaks in infrared and eventually in radio, marks the disappearance of the hot inner part of an accretion disk crossing the horizon of the central BH.
As the disk matter is continuously replenished by the companion star, the disk has to evacuate the subsequent excess of rotational kinetic energy density, triggering the launch of the bipolar jets.
The emitted plasma clouds expand while radiating, reducing the opacity to their own radiation and freeing the optical path first to IR and then to Radio photons.
This is the best documented case, but not the only one, grounding the relation between plasma disks dynamics and radio-jets genesis.

\paragraph{\bf The Shakura Standard Model --}
The Standard Model for accretion disks allows the inward matter flux preserving the balance of total angular momentum because of viscous stresses, explicitly parametrized by a coefficient $\alpha$ which scales the viscosity coefficient $\visco$  \cite{S73}.
Although a non-vanishing viscosity exists in astrophysical plasmas also in quasi-ideal conditions, its value is too much low to reach the accretion rate values provided by observations.
To deal with this problem, the Standard Model introduces a \emph{turbulent} enhancement of the viscosity (and of the $\alpha$ coefficient) which accounts for about 8 orders of magnitude over the ideal (laminar) value, as estimated for ion-ion collisions when the central object is a NS with $\Rstar \simeq 10 \km$ surrounded by a disk with height $H \simeq 10^3 \km$, temperature $T\simeq 10^6 \kelvin$ and density $\ne \simeq 10^{10} \cm^{-3}$, accreting with an efficiency $\sim 10^{-3}$.

The currently most believed mechanism addressed to trigger the turbulence (and sustain it) is a linear MHD instability generated by the coupling of a weak magnetic field to an outwardly decreasing angular velocity profile.
This has been originally formalized as an instability for magnetized fluids arranged in a cylindrical Couette flow  \cite{V59} and then extended to the most general rotational case  \cite{CHANDRA60}, but it needed three decades to settle in and unveil its relevance for accretion disks  \cite{BH91}.
Such Magneto-Rotational Instability (MRI) claims that weak magnetic fields actively generate turbulence, instead of passively being advected/disrupted by it as it is found in the classical Shakura description  \cite{BH98REVIEW}.
The key to the phenomenon is in the magnetic tension threading two contiguous layers of differentially rotating plasma, which acts on them like a spring between two mass points orbiting at close radii: a weak spring is able to transfer momentum from the inner to the outer one, forcing the first to drop down to an even inner orbit and the second to move outwards, so that the separation between them grows exponentially.
The overall outcome of the process is to break coherent fluid motions, producing significant Maxwell stresses and enhancing the angular momentum transport; the net effect has been quantified via simulations of fully-developed turbulent motion, which lead to estimations  \cite{hawley95,brandenburg95} of the effective viscosity in terms of $\alpha \simeq 0.005 \div 0.5$, depending on the initial topology of the field.
It is worth noting that the stability criterion doesn't depend on the geometry of the field but only on its strength and on the plasma mean density and temperature.
Any configuration seems to evolve into a turbulent state with growing field, up to a saturated state which depends on a competitive process not yet fully understood.
If MRI is efficient in preserving turbulence at the field saturation, the resulting highly agitated flow can give a critical feedback on the magnetic field because of the stretching of the field lines, possibly resulting in the amplification usually known as dynamo effect  \cite{fromang07}.
A variety of papers has then studied and characterized the MRI-driven turbulence.
Without going beyond the scope of this work, we suggest to the interested reader the most recent studies about: the effect of non-axisymmetric perturbations  \cite{bodo13}; the interplay with magneto-centrifugal jet launching  \cite{fromang13}; the global structure of an MRI-turbulent disk  \cite{beckwith11}.

Nevertheless, some systems exist that are not satisfactory described -- whose spectra are not correctly fitted -- for any choice of the enhanced viscosity (i.e., for any value of $\alpha$) nor of any other Standard Model parameter (e.g., RW Sextantis \cite{linnell10}, a weakly magnetized WD).
This problem stands outside the discussion about the nature of the turbulence and the reliability of the MRI.
Since the $\alpha$-prescription affects the accretion rate because $\dot{m} = \dot{m}\<{\visco\<{\alpha}}$, from the definition $\dot{m} = 2 \pi r v_r \rho H$ we can derive the radial velocity $v_r = \<{{H}/{r}} \alpha \vS$ in terms of the sound speed $\vS$.
An equivalent expression should be obtained from the Generalized Ohm Law which states:
\SUBEQ{
\begin{equation} \label{GOL}
\mathbf{E} + \dfrac{1}{c} (\mathbf{v} \times \mathbf{B} ) = \dfrac{1}{\conduco} \mathbf{J} \, ,
\end{equation}
\begin{equation}
\dfrac{v_r B_0}{c} \simeq \dfrac{c}{4\pi \conduco} \dfrac{B_1}{\lambda} \, ,
\end{equation}
}
where $\conduco$ is the conductivity coefficient, $B_{0,1}$ is the magnitude of background and back-reaction field respectively, and $\lambda$ is the back-reaction length scale (note that $B_1$ and $\lambda$ are responsible for the currents $J_\phi$ induced in the disk); the latter equation is simply derived from the azimuthal component of the former.
Equating the expressions for $v_r$, we are able to provide an estimation for the accreting plasma Magnetic Prandtl Number (PrM), which quantifies the relevance of the viscous effects over the resistive ones:
\begin{equation} \label{prandtl}
\mathrm{PrM} \<{n,T} \doteq \dfrac{4 \pi \visco \conduco}{c^2 \rho}
\simeq \dfrac{R_\mathrm{in}}{\lambda} \dfrac{B_1}{B_0} \, ,
\end{equation}
with $R_\mathrm{in}$ as the inner boundary of the disk.
This leads us to two observations, considering that the kinetic estimations easily give $\mathrm{PrM} \gg 1$ in our range of interest, namely temperature in $10^5 \div 10^7 \kelvin$ and particles density in $10^8 \div 10^{12} \cm^{-3}$.
First, since the Standard Model neglects back-reaction and azimuthal currents, it asks for $B_1 \ll B_0$ and $\lambda \simeq R_\mathrm{in}$: this implies an \virg{effective} PrM which is small at best \cite{BKL01}.
Since the adopted viscosity is already turbulence-enhanced by a $10^8$ factor, the Shakura Model needs a surprisingly small (anomalous) conductivity \cite{MC12}.
On the other hand, for reasonable fields $B_1 \lesssim B_0$, we have to claim that $\lambda \ll R_\mathrm{in}$ to be consistent with the high values of the quasi-ideal PrM: a realistic (i.e., non-effective) model lies towards a slightly different approach which embodies the formation of magnetic microstructures  \cite{MP13}.

\begin{figure} 
\centering
\FIGURE[.48]{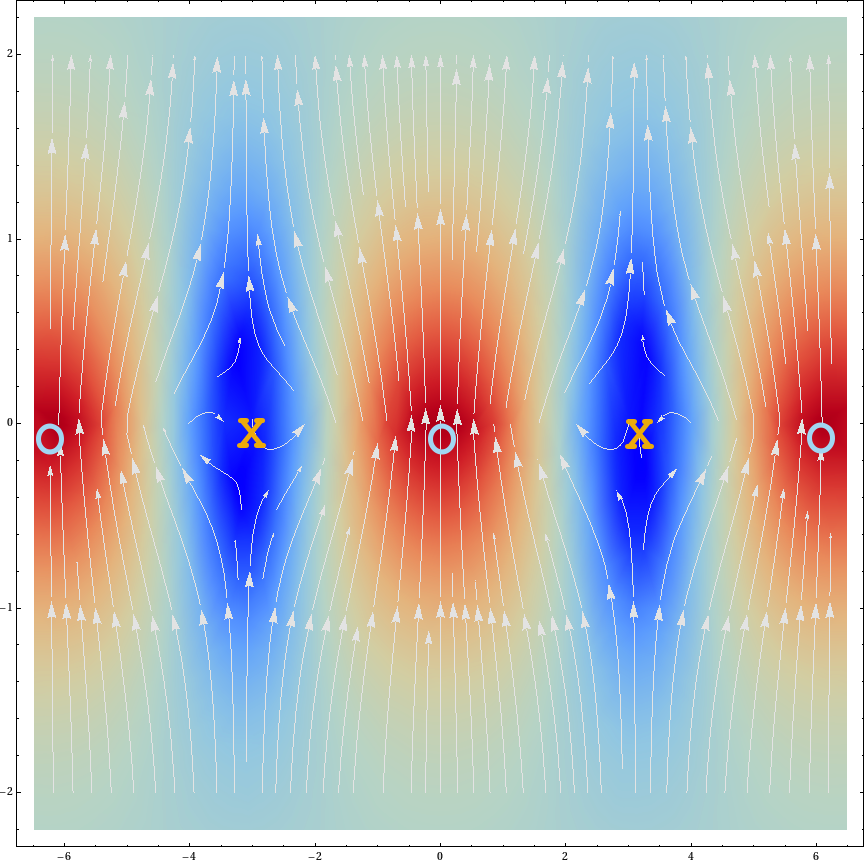}
\FIGURE[.46]{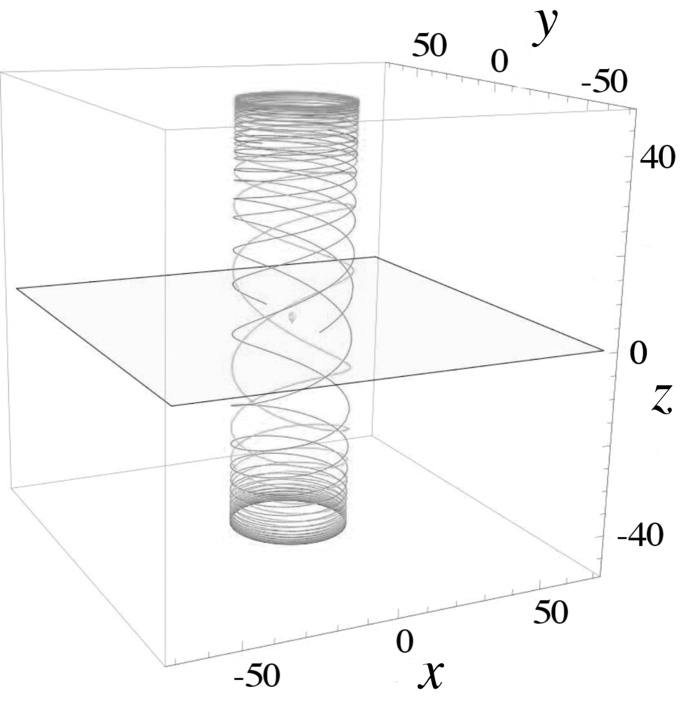}
\caption{ \label{fig:B}
(Color online)
\emph{Left:}
Plot on the local poloidal plane of dimensionless radius and height -- units are arbitrary and inessential; colors are for the values of the magnetic flux surfaces (darker shades for greater magnitude), arrows are for the actual magnetic field, as expressed in Equation \eqref{Bfield}. The yellow Xs mark the X-points of zero $B_z$, while the cyan Os mark the O-points of zero $B_r$.
\emph{Right:}
Three-dimensional view in Cartesian coordinates of two jet streamlines, twisted together by the rotation of the accretion disk, whose equatorial height is suggested by the shaded plane. Only two different plasma trajectories are represented, placed at a fixed radius and at the azimuthal angles $\varphi=0,\pi$ respectively. It is shown the funnel-like nature of the axial jet seed and the high collimation of the stream, which is however confined inside the disk in quasi-ideal conditions: the induced magnetic field lines open to outer space only if a finite conductivity is adopted.
Adapted from Figure 4 in Tirabassi, Montani \& Carlevaro \cite{MCT13}; reproduced by permission of the American Physical Society.
}
\end{figure}

\paragraph{\bf The microstructures paradigm --}
In the pioneering work of Coppi \& Rousseau \cite{CR06}, an alternative model for a plasma accretion disk has been depicted.
It deals with a quasi-ideal plasma, whereas the Standard Model owns a neutral viscous fluid, and retains the vertical equation which was previously averaged out by the Shakura one-dimensional approach.
This way we get a two-dimensional coupled dynamics richer than the traditional one, where every relevant physical quantity only depends on the radial coordinate.
Via a local analysis in the neighbourhood of a fixed radius $r=R_0$, it has been shown the formation of a small-scale, periodic, rigid (i.e., non-diffusive) structure of the magnetic flux surfaces, corresponding to a back-reaction magnetic field critically affecting the equilibrium configuration.
The total magnetic field assumes in this case the following expression:
\begin{align} 
B_r & = \Psi_{1,0} \dfrac{z}{R_0 H_0^2} \exp\<{- \dfrac{z^2}{H_0^2}} \sin\<{k_0\<{r-R_0}} \nonumber \\
\label{Bfield} \\
B_z & = B_{0z} + \Psi_{1,0} \dfrac{k_0}{R_0} \exp\<{- \dfrac{z^2}{H_0^2}} \cos\<{k_0\<{r-R_0}} \, , \nonumber
\end{align}
where $r,z$ are the radial and vertical coordinate respectively, $H_0$ is the disk height at the fiducial radius $R_0$, $k_0 \doteq 2\pi / \lambda$ determines the back-reaction length scale and $\Psi_{1,0}$ is a real constant related to the magnitude of the induced field (a visual representation is in \Fref{fig:B}, top panel).
If the induced inner currents are strong enough, the magnetic structure may lead to a corrugated density profile and eventually to the fragmentation of the disk into a ring sequence.

It is worth noting that the solution \eqref{Bfield} works in Equation \eqref{GOL} to give substantial hints on brand new ways to produce accretion and trigger jets.
First, the Generalized Ohm Law allows radial infall because of the X-points where the vertical magnetic field vanishes \cite{MB11}, since it can be recasted to say $v_r \simeq {v_z B_r}/{B_z}$, which arbitrarily grows where $B_z \rightarrow 0$.
This picture clearly needs another non-steady mechanism able to push the stationary flow toward and through the X-points, and the major candidate is thought to be a modified version of the \virg{ballooning modes}, well known in the context of laboratory plasma physics \cite{C03} but not yet applied to the astrophysical setting.
On the other hand, the same Law allows vertical ejection because of the O-points where the radial magnetic field vanishes, since it can be conversely rewritten as $v_z \simeq {v_r B_z}/{B_r}$, stating that the local vertical velocity diverges where $B_r \rightarrow 0$.
This proportionality statements are more striking when the plasma is assumed to be ideal, so there is no current contribution in the Equation \eqref{GOL} acting as a smoothing diffusive term.
But in this case, the velocity field results parallel to the magnetic field.
The motion of the jet seed is therefore closed since the particle trajectories are frozen to the magnetic flux surfaces (see \Fref{fig:B}, bottom panel).
An actual matter flux outward the disk needs some misalignment of $\mathbf{v}$ and $\mathbf{B}$, offered by dissipative effects: a finite conductivity $\conduco$ opens the plasma streamlines allowing the effective plasma outflow through the jet \cite{MCT13}.

\paragraph{\bf Jet triggering --}
The paradigm described in the previous paragraph is different from the most believed picture of jet formation, which invokes differential rotation of the poloidal component of the magnetic field either in the inner disk or in the BH ergosphere  \cite{meier01}.
In the case of a Standard-disk-propelled jet, it seems hard to reach ultrarelativistic flow speeds because of the diffusive nature of the magnetic field in the accepted Shakura Model: the material ejected at the local escape speed cannot find a source of magnetic energy useful to reach the observed jet speed.
On the other hand, the jets launched from the ergosphere of a rotating BH can be described in the so-called Blandford-Znajek (BZ) model  \cite{BZoriginal}, which is endowed with observational and numerical support.
These BZ jets are powered by the extraction of rotational energy from the BH.
This is possible because of the presence of the disk magnetic field, rotating at a speed $\Omega_F$ and threading the event horizon, which encloses a spacetime rotating at a different speed $\Omega_H$.
When the field strength is large enough, a force-free magnetosphere can be established in a funnel-shaped region around the BH rotational axis.
Here the vacuum is unstable to $e^+e^-$ cascades, able to drain the BH of a power:
\begin{equation} \label{BZpower}
P_\mathrm{BZ} \propto \dfrac{\Omega_F}{\Omega_H} \<{1-\dfrac{\Omega_F}{\Omega_H}} \Phi^2 a^2 \, ,
\end{equation}
depending on the rotational frequencies (with maximum efficiency at the resonance $\Omega_H = 2 \Omega_F$), on the magnetic flux threading the jet $\Phi$ and on the dimensionless spin parameter of the BH $a$.
The Lorentz factor of such a jet can be arbitrarily large \cite{BZmckinney} (i.e., the jet can be arbitrarily fast) because of the high initial energy-to-rest-mass ratio provided: this jet is made up of free electrons generated on the poles above the event horizon, while any other mechanism gives jets loaded directly by disk matter.
Observational evidences that jets may be powered by BH spin energy exist  \cite{BZnarayan, BZmcclintock}, as it has been proved that for impulsive ballistic jets (emitted by BH transient systems via outbursts) the peak radio luminosity is a reliable proxy for the jet kinetic energy, and the jet power is in good agreement with the BZ model predictions.
The same confirmation has come from General Relativistic MHD simulations  \cite{BZpenna} of jets ejection by spinning BHs (up to $a \simeq 0.98$) accreting from geometrically thick disks with $H / R \sim 0.3$.
On the different scale of AGNs, a flux-trapping variant  \cite{BZgarofalo} of the BZ model has been carried out via GR-MHD simulations of a thin disk in Kerr spacetime, showing a \virg{plunge} region of free fall inside the radius of marginal stability of the disk.
The size of this region (and the efficiency of the flux-trapping) depends on the \emph{signed} value of the BH spin, deviating from \Eref{BZpower} and suggesting that the most powerful radio galaxies needs \emph{retrograde} rapidly rotating BHs, which is compatible at least with observations  \cite{BZkataoka} of the galaxy 3C 120.

It is worth noting that the link between the Angular Momentum Transport, 
expected to be at the ground of the jet formation, 
and the accretion features can be outlined in both Quasar and Microquasar sources.
In particular, the variation of the Neutron Star spin in a Microquasar is clearly induced by the accretion of material from the companion star, as far as they are compared with the regular spin-down behavior of  isolated Pulsars. 
On the other hand, the Quasar power emission can be properly accounted for only if they 
are interpreted as accretion-powered structures, 
so getting acceptable mass accretion rates. 
Thus, both these types of sources 
are characterized by an intense radio emission, 
accretion of matter on a compact central object and 
the presence of a marked highly-collimated jet. 
We cannot regard these sources as 
the smoking gun of the magnetic microstructure 
paradigm -- which requires in itself a significant effort to 
account for the detected accretion rates -- but the very different scale of Quasars and Microquasars suggests that the explanation for jet formation has reliably to do with the 
accretion morphology of the sources, more than with peculiar properties of the corresponding 
central bodies. 
In this sense, we must look at them as the natural arena in which implement and 
test the reformulation of the 
basic paradigm underlying the Angular Momentum Transport. 

\section{Concluding Remarks} \label{sec:conclusions}
Our review of the radio-frequency sources in astrophysics has outlined how the emission in the radio band is very relevant in characterizing and classifying important classes of stellar and galactic objects,
giving to the radio astronomy the status of a specific discipline.
A crucial role in this context is played by the jet emission along the axis of highly energetic sources, like Microquasars, Quasars and Radio-Loud Galaxies, especially when the constituent plasma of this very collimated structures cools enough that the synchrotron emission peaks in the radio band.
We have highlighted the universality of this feature, and we investigated the related problems of jet triggering and jet-disk coupling, which enforce the link between accretion phenomena and jet ejection.
Moreover, they both deal with basic plasma physics.

We also clarified how the huge amount of radio-energy (up to $10^{40}$ W in Quasars and Blazars) is essentially due to the accretion power onto compact objects, but its real origin is not yet well-understood.
In particular, we outlined how the standard Shakura picture 
for the accretion mechanism requires very large values 
of the plasma viscosity and resistivity, which 
are not justified on a fundamental point 
of view and therefore appear as a fine-tuning of 
the model.   
As an alternative perspective, we then inferred 
the emergence of magnetic microstructures 
in the quasi-ideal plasma, which can 
be responsible for plasma porosity effects nearby the 
X-points of the resulting magnetic profile 
(indeed the magnetic surfaces acquire a 
small-scale radial oscillation). 
The existence of such a crystal-like 
profile of the magnetic field offers also 
a very favourable mechanism for jet-seed generation in 
the corresponding O-points, where the radial magnetic field 
component vanishes. In fact, the azimuthal component 
of the electron force balance determines peaks 
of the vertical velocity nearby these O-points, 
so that the collimated nature of the jets around a 
given value of the radial coordinate is naturally 
guaranteed.  

Thus, observing astrophysical radio sources and their morphology, we can recognize that the problem of Angular Momentum Transport across the accreting structures, which is certainly at the ground level of the accretion dynamics (see the Microquasar features), is far from being consistently described.
It is just in this open question that the cross-fertilization among astrophysical and laboratory plasma must be enforced in order to trace the common plasma physics paradigm.
In fact, both the accretion disks and the Tokamak configurations are axially symmetric equilibria, both are concerned with a rotation profile and, eventually, the radio-energy interacting with, or generated by, these two plasmas has a relevant impact on their quasi-steady state and its stability.
Furthermore in both these two systems, the Angular Momentum Transport is a crucial phenomenon in fixing the nature of the plasma magnetic confinement.

\end{document}